\documentclass[aps,twocolumn,groupedaddress,superscriptaddress,showpacs]{revtex4-1}

\usepackage{graphicx}
\usepackage{color}
\usepackage{amsmath}
\usepackage{amsfonts}
\usepackage{amssymb}
\usepackage{dcolumn}
\usepackage{float}

\begin{document}
  \title{Towards designing robust coupled networks}

  \author{Christian M. Schneider}
    \affiliation{Department of Civil and Environmental Engineering,
Massachusetts Institute of Technology, 77 Massachusetts Avenue,
Cambridge, MA 02139, USA}

  \author{Nuri Yazdani}
    \affiliation{Computational Physics for Engineering Materials, IfB,
ETH Zurich, Wolfgang-Pauli-Strasse 27, CH-8093 Zurich, Switzerland}

  \author{Nuno A. M. Ara\'ujo}
    \email{Correspondence and requests for materials should be addressed to N. A. M. A. (nuno@ethz.ch)}
    \affiliation{Computational Physics for Engineering Materials, IfB,
ETH Zurich, Wolfgang-Pauli-Strasse 27, CH-8093 Zurich, Switzerland}

  \author{Shlomo Havlin}
    \affiliation{Department of Physics, Bar-Ilan
University, 52900 Ramat-Gan, Israel}

  \author{Hans J. Herrmann}
    \affiliation{Computational Physics for Engineering Materials, IfB,
ETH Zurich, Wolfgang-Pauli-Strasse 27, CH-8093 Zurich, Switzerland}
    \affiliation{Departamento de F\'isica, Universidade Federal do
Cear\'a, 60451-970 Fortaleza, Cear\'a, Brazil}


\begin{abstract}
Natural and technological interdependent systems have been shown to be highly
vulnerable due to cascading failures and an abrupt collapse of global connectivity under
initial failure.  Mitigating the risk by partial disconnection endangers their
functionality. Here we propose a systematic strategy of selecting a
minimum number of autonomous nodes that guarantee a smooth transition in
robustness. Our method which is based on betweenness is tested on
various examples including the famous $2003$ electrical blackout of
Italy. We show that, with this strategy, the necessary number of
autonomous nodes can be reduced by a factor of five compared to a
random choice.  We also find that the transition to abrupt collapse
follows tricritical scaling characterized by a set of exponents which is
independent on the protection strategy. 
\end{abstract}

  \maketitle

Interconnected complex networks are ubiquitous in today’s world. They
control infrastructures of modern society
(energy-communication-transportation), the financial system or even the
human body \cite{Peerenboom01,Rosato08,Schweitzer09}. Unfortunately, they
are much more fragile than uncoupled networks as recently recognized
through the finding that the robustness changes from a second order
transition in uncoupled systems to first order in interdependent systems
\cite{Buldyrev10,Brummitt12,Gao2012}. The
obvious mitigation strategy consists in partially decoupling
the networks by the creation of autonomous nodes \cite{Parshani10}. Too
much disconnection however risks endangering the functionality of the
system. The question which we will address here is how to
reduce fragility without losing functionality and we  will in fact
answer this question by developing an explicit algorithm based on
betweenness that enables to avoid the abrupt collapse with a minimum number of
autonomous nodes.

  \begin{figure}
    \includegraphics[width=\columnwidth]{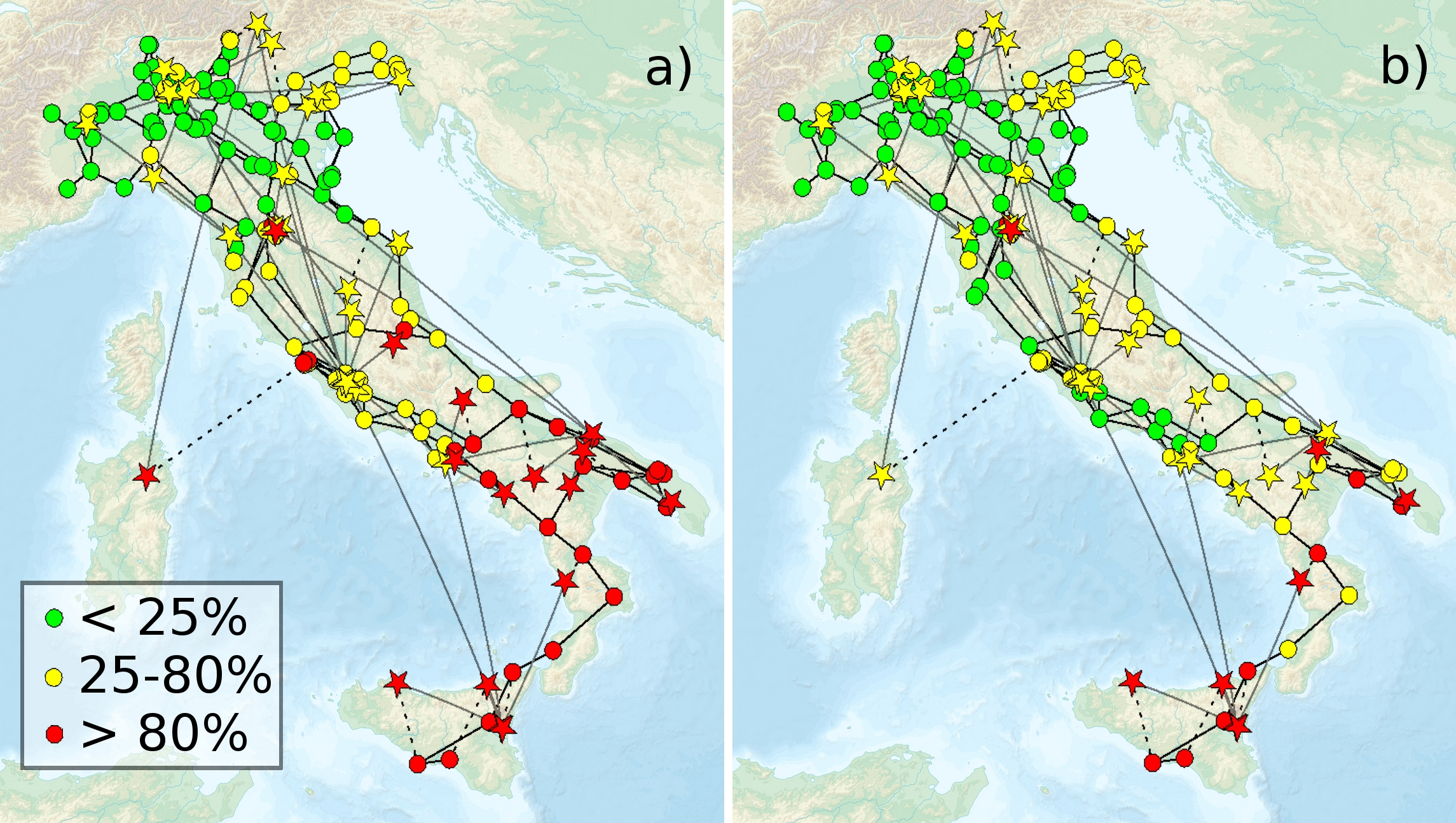}
    \caption{
      The herein proposed strategy improves significantly
the resilience of the coupling between the communication system (39
stars) and the power grid (310 circles) in Italy. The color scheme
stands for the probability that the node is inactive after the random
failure of $14$ communication servers. In a) all communication servers
are coupled while in b) four servers have been decoupled following the
strategy proposed here. The coupling between the networks was
established based on the geographical location of the nodes, such that
each communication server is coupled with the closest power station
\cite{Rosato08}. The images were produced using the software Pajek.
\label{fig::italy}
    }
  \end{figure}
  Buldyrev {\it et al.} \cite{Buldyrev10} proposed a percolation
framework to study two coupled networks, $A$ and $B$, where each
$A$-node is coupled to a $B$-node, via bi-directional links, such that
when one fails the other cannot function either. The removal of a
fraction of $A$-nodes may trigger a domino effect where, not only their
counterparts in $B$ fail, but all nodes that become disconnected from
the giant cluster of both networks also fail. This causes further
cascading of failures, yielding an abrupt collapse of connectivity,
characterized by a discontinuous (first order) percolation transition.
Parshani {\it et al.} \cite{Parshani10} showed that damage can be
mitigated by decreasing the degree of coupling, but only if a
significant fraction ($\approx 0.4$) of nodes is decoupled, the
transition changes from discontinuous to continuous. The coupling is
reduced by randomly selecting a fraction of nodes to become autonomous
and, therefore, independent on the other network. For the coupling
between power stations and communication servers, for example,
autonomous power stations have alternative communication systems which
are used when the server fails and an autonomous server has its own
energy power supply. We propose a method, based on degree and
centrality, to identify these autonomous nodes that maximize the
robustness of the system in terms of connectivity.  We show that, with this scheme,
the critical coupling increases, i.e., the fraction of nodes that needs
to be decoupled to smoothen out the transition is much smaller (close to
$0.1$ compared to $0.4$).  Significant improvement is observed for
different coupled networks including for Erd\H{o}s-R\'enyi graphs (ER)
where such improvement in the robustness was unexpected given their
narrow degree distribution. To demonstrate the strength of our approach,
in Fig.~\ref{fig::italy} we apply the proposed strategy to the real
coupled system in Italy \cite{Rosato08} and show that by only protecting
four servers the robustness is significantly improved (details in the
figure caption).  

  \begin{figure}
    \includegraphics[width=\columnwidth]{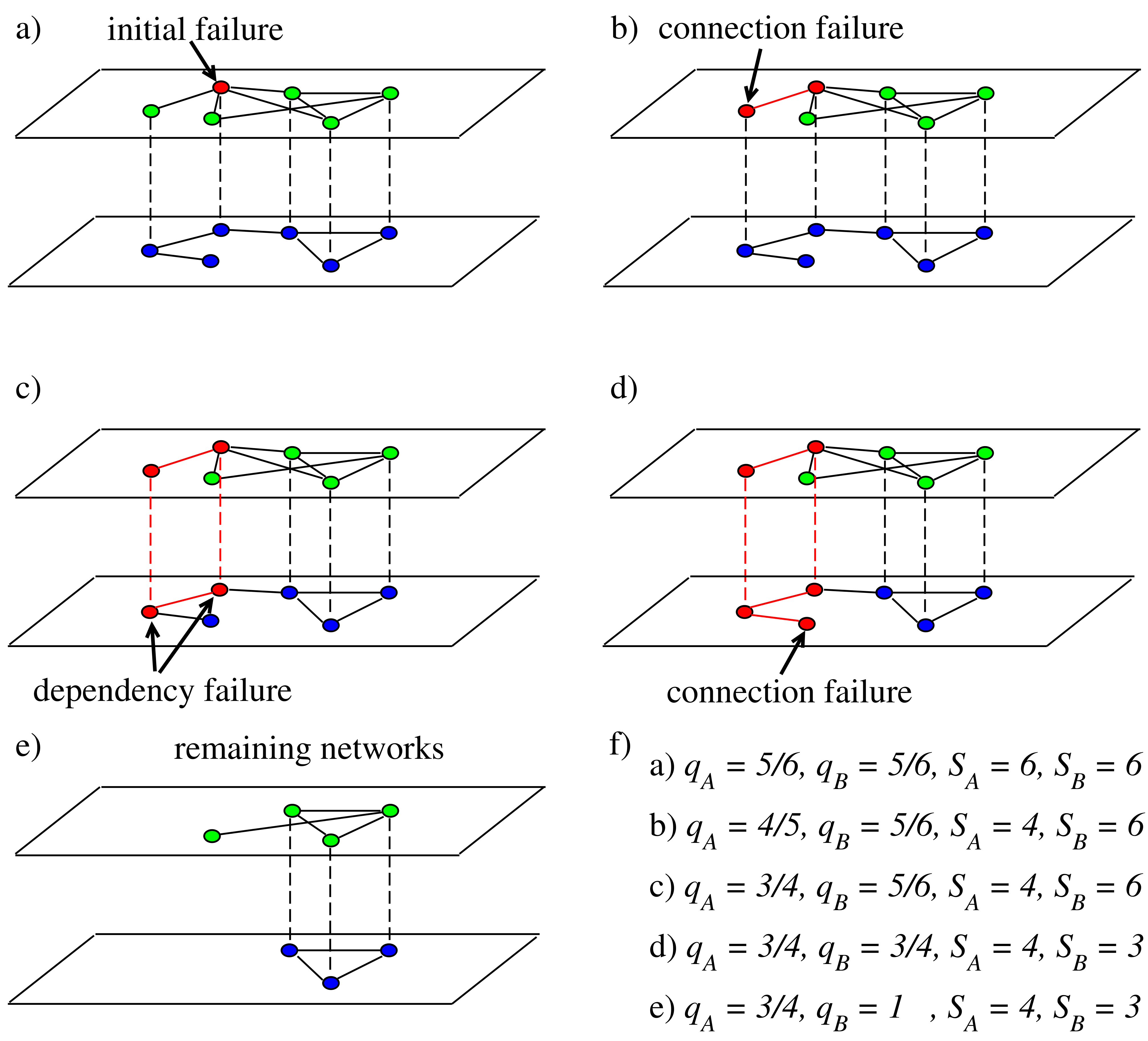}
    \caption{Scheme of the cascade of node failures triggered by the
initial failure of a node in network $A$ (top network). Two networks, $A$
(top) and $B$ (bottom), are considered. When a node initially fails in
network $A$ (a) all nodes connected to the largest component through it
also fail (b) as well as the corresponding dependent nodes in network $B$
(c). The failure of the dependent nodes in network $B$ leads to further
failures in both networks (d) and (e). For each iteration step, the
degree of coupling $q_x$ and the size of the largest connected component
$S_x$ for each network $x$ are listed in (f).
       \label{fig::scheme}
    }
  \end{figure}
  We consider a pair of networks, $A$ and $B$, where a fraction $q$
(degree of coupling) of $A$-nodes are coupled with $B$-nodes. To be
functional, nodes need to be connected to the giant cluster of their
network. When an $A$-node fails, the corresponding one in $B$ cannot
function either. Consequently, all nodes bridged to the largest cluster
through these nodes, together with their counterpart in the other
network, become also deactivated. A cascade of failures occurs with
drastic effects on the global connectivity (see Fig.~\ref{fig::scheme}) \cite{Buldyrev10,Parshani10}.
This process can also be treated as an epidemic spreading \cite{Son11b}.
To study the resilience to
failures, we follow the size of the largest connected cluster of active
$A$-nodes, under a sequence of random irreversible attacks to network
$A$.  Notwithstanding the simplicity of solely considering random
attacks, this model can be straightforwardly extended to targeted ones
\cite{Huang10}. 
Recently, for single networks, it has been proposed \cite{Schneider11} to
quantify the robustness $R$ as
  \begin{equation}\label{eq::def.robustness}
    R=\frac{1}{N}\sum_{Q=1}^{N}S(Q) \ \ ,
  \end{equation}
where $Q$ is the number of node failures, $S(Q)$ the size of the largest
connected cluster in a network after $Q$ failures, and $N$ is the total
number of nodes in the network \cite{Schneider11,Herrmann10}. 
Here we extend this
definition to coupled systems by performing the same measurement, given
by Eq.~(\ref{eq::def.robustness}), only on the network where the random
failures occur, namely, network $A$.
To follow the cascade triggered by the
failure of a fraction $1-p$ of $A$-nodes, similar to \cite{Parshani10},
we solve the iterative equations,
\begin{eqnarray}\label{eq::iterative}
  \beta_n &=& 1-q_{\beta,n}\left[1-S_{A}\left(\alpha_n\right)p\right]
, \\
  \alpha_n &=& p\left(1-q_{\alpha,n}\left[1-S_{B}\left(\beta_{n-1}\right)\right]\right) ,
\end{eqnarray}
with the initial condition $\alpha_1=p$, where $\alpha_n$ and $\beta_n$
are the fraction of $A$ and $B$ surviving nodes at iteration step $n$
and $S_{x}(\chi_n)$ is the fraction of such nodes in the giant
cluster. $q_{\chi,n}$ is the fraction of dependent nodes in network
$\chi$ fragmented from the largest cluster (see {\it Methods} for further details).

\section*{Results}

    \begin{figure}
      \includegraphics[width=0.7\columnwidth]{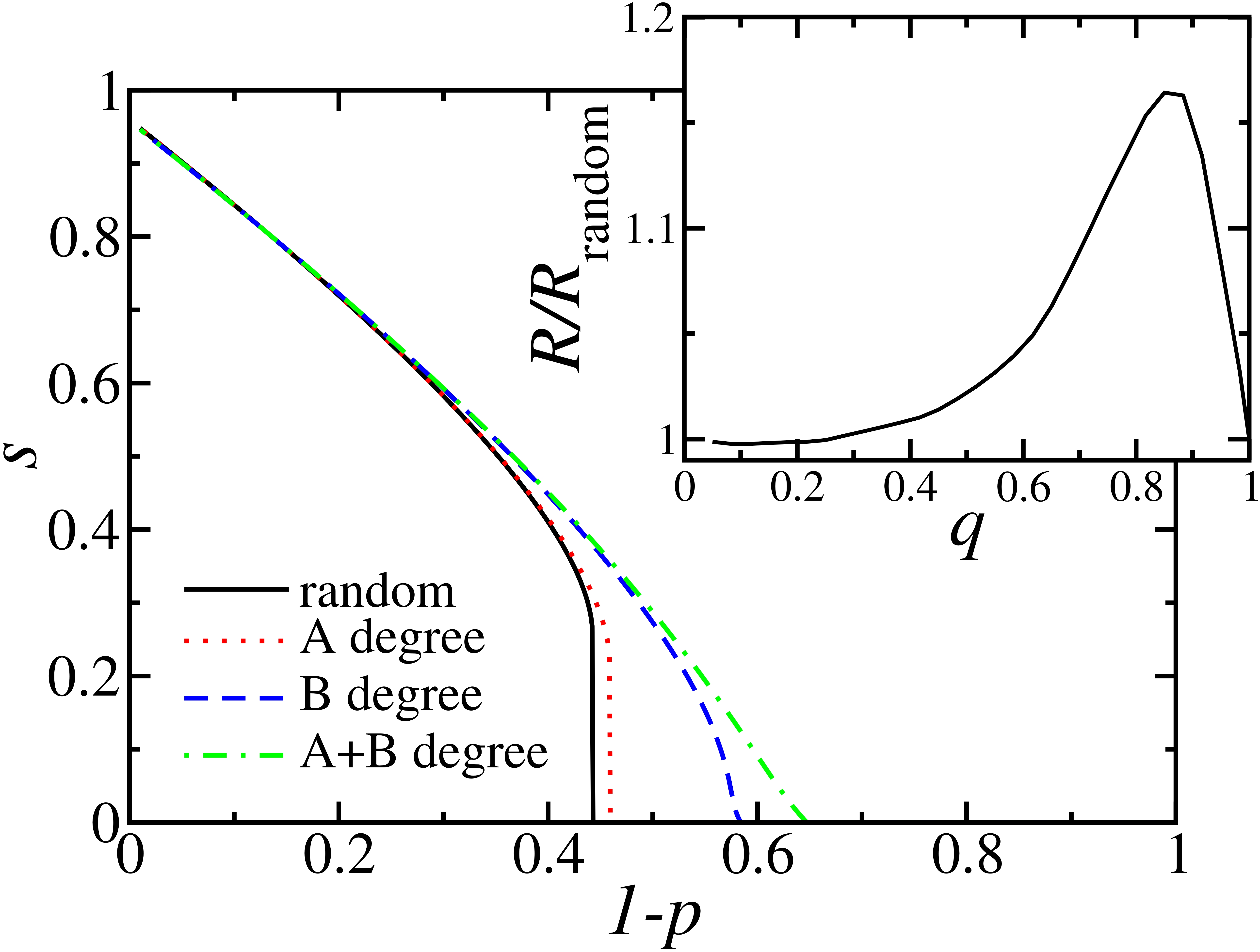} 
      \caption
      { \label{fig::orderparam}
	Fraction of $A$-nodes in the largest connected
cluster, $s$, as a function of the fraction of randomly removed nodes
$1-p$ from network A, for two coupled ER (average degree $\langle
k\rangle =4$) with $90\%$ of the nodes
connected by inter-network links ($q=0.9$). It is seen that robustness can
significantly be improved by properly selecting the autonomous nodes.
We start with two fully interconnected ER and decouple $10\%$ of their
nodes according to three strategies: randomly (\mbox{(black-)solid}
line), the ones with highest degree in network $A$ (\mbox{(red-)dotted}
line) and in network $B$ (\mbox{(blue-)dashed} line).  We also include
the case where $10\%$ autonomous nodes in both networks are chosen as the ones with
highest degree and all the others are interconnected randomly
(\mbox{(green-)dotted-dashed} line).  The inset shows the dependence of
the relative robustness of the degree strategy on the degree of coupling
$q$ compared with the random case.  Results for the degree have been obtained
with the formalism of generation functions (see {\it Methods}).
      }
    \end{figure}
To demonstrate our method of selecting autonomous nodes we consider two
ER graphs with average degree $\langle k\rangle = 4$ and $10\%$ of
autonomous nodes ($q=0.1$). First
we consider a method based on the degree of the node and later we
compare with the method based on the betweenness.
Under a sequence of random failures, the networks are catastrophically
fragmented when close to $45\%$ of the nodes fail, as seen in
Fig.~\ref{fig::orderparam}.  For a single ER, with the same average
degree, the global connectivity is only lost after the failure of $75\%$
of the nodes.  Figure~\ref{fig::orderparam} also shows
(\mbox{(green-)dotted-dashed} line) the results for choosing as autonomous nodes
in both networks the fraction $1-q$ of the nodes with the highest degree and
coupling the remaining ones at random. With this strategy,
the robustness $R$ can be improved and the corresponding
increase of $p_c$ is about $40\%$, from close to $0.45$ to close to
$0.65$. Also the order of the transition changes from first to second
order. Further improvement can be achieved if additionally the coupled nodes
are paired according to their position in the ranking of degree, since
interconnecting similar nodes increases the global robustness
\cite{Parshani10c,Buldyrev11}. In the inset of
Fig.~\ref{fig::orderparam} we see the dependence on $q$ of the relative
robustness for the degree strategy compared to the random case
$R/R_\mathrm{random}$. For the entire range of $q$ the proposed strategy
is more efficient and a relative improvement of more than $15\%$ is
observed when still $85\%$ of the nodes are coupled.

  Two types of technological challenges are at hand: either a system has
to be designed robust from scratch or it already exists, constrained to
a certain topology, but requires improved robustness.  In the former
case, the best procedure is to choose as autonomous the nodes with
highest degree in each network and couple the others based on their rank
of degree.  For the latter, rewiring is usually a time-consuming and
expensive process, and the creation of new autonomous nodes may be
economically more feasible.  The simplest procedure consists in choosing as
autonomous both nodes connected by the same inter-network link.
However, a high degree node in network $A$ is not
necessarily connected with a high degree node in network $B$.
 In Fig.~\ref{fig::orderparam} we compare between choosing the autonomous
pairs based on the degree of the node in network $A$ or in network $B$.
When pairs of nodes are picked based on their rank in the network under
the initial failure (network $A$), the robustness almost does not
improve compared to choosing randomly.  If, on the other hand, network
$B$ is considered, the robustness is significantly improved, revealing
that this scheme is more efficient.  This asymmetry between $A$ and $B$
network is due to the fact that we attack only nodes in network $A$,
triggering the cascade, that initially shuts down the corresponding
$B$-node.  The degree of this $B$-node is related to the number of nodes
which become disconnected from the main cluster and consequently affect
back the network $A$.  Therefore, the control of the degree of
vulnerable $B$-nodes is a key mechanism to downsize the cascade.  On the
other hand, when a hub is protected in network $A$ it can still be
attacked since the initial attack does not distinguish between
autonomous and non-autonomous nodes.

    \begin{figure}
      \includegraphics[width=0.7\columnwidth]{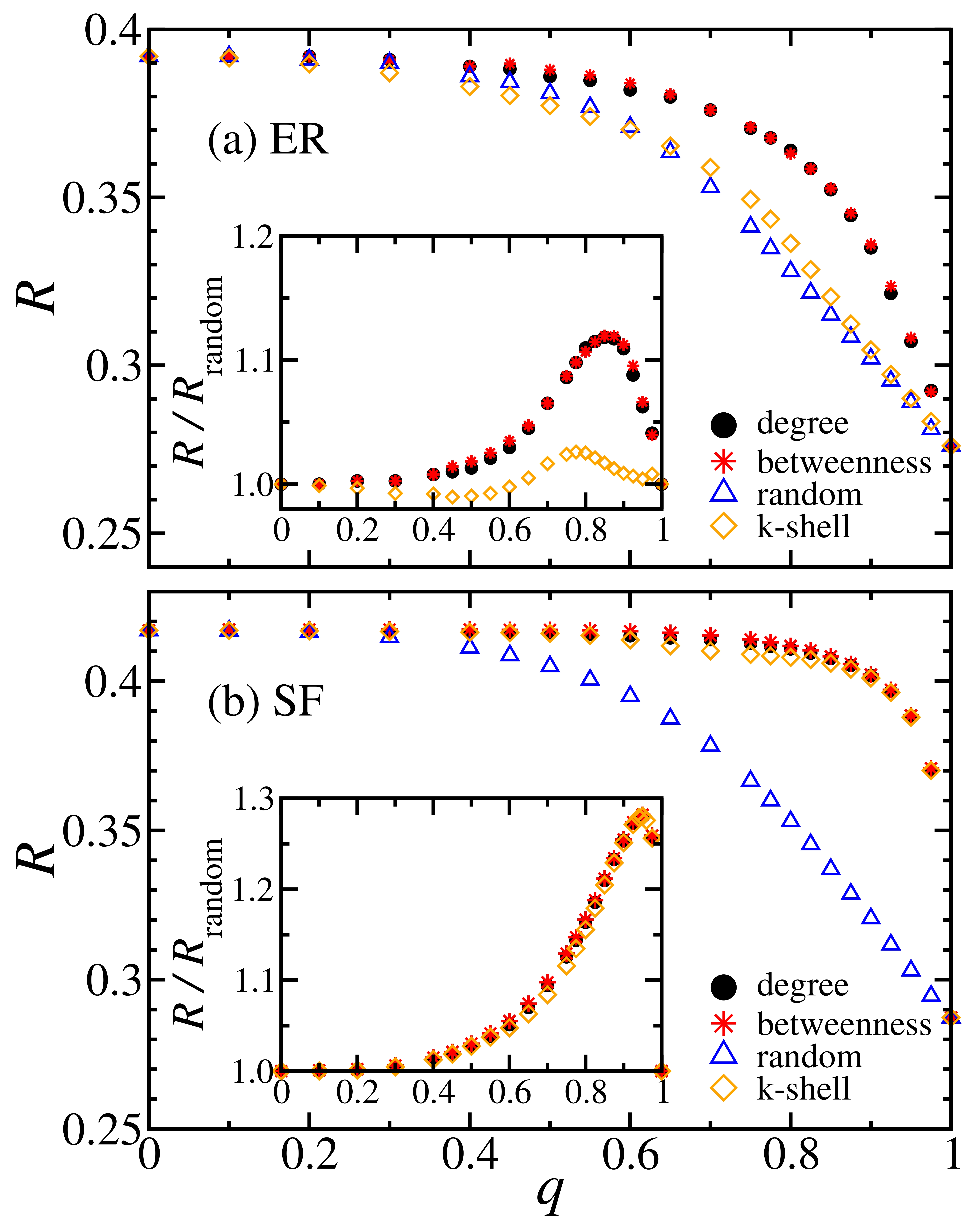} \\
      \caption
      { \label{fig::robustness}
	Dependence of the robustness, $R$, on the degree
of coupling, $q$, for two, interconnected, (a) ER (average degree
$\langle k\rangle = 4$)
and (b) SF with degree exponent $\gamma=2.5$. Applying our 
proposed strategy is applied, the optimal fraction of autonomous nodes
is relatively very small.  Autonomous nodes are chosen in four different
ways: randomly (\mbox{(blue-)triangles}), high degree
(\mbox{(black-)dots}), high betweenness (\mbox{(red-)stars}), and high
k-shell (\mbox{(yellow-)rhombi}).  The insets show the relative
improvement of the robustness, for the different strategies of
autonomous selection compared with the random case.  Results have been
averaged over $10^2$ configurations of two networks with $10^3$ nodes
each.  For each configuration we averaged over $10^3$ sequences of
random attacks.
      }
    \end{figure}
  In Fig.~\ref{fig::robustness}(a) we compare four different criteria to
select the autonomous nodes: betweenness, degree, k-shell, and random
choice, for two coupled ER networks. In the betweenness strategy, the
selected autonomous are the ones with highest betweenness. The
betweenness is defined as the number of shortest paths between all pairs
of nodes passing through the node \cite{Newman10}. 
A k-shell is obtained by removing, iteratively, all nodes with degree
smaller than $k$, until all remaining nodes have degree $k$ or larger.
In the k-shell
strategy, the autonomous are chosen as the ones with highest k-shell in the
k-shell decomposition \cite{Carmi07}. 
The coupled nodes (not autonomous), for all
cases, have been randomly inter-linked.  Since ER networks are
characterized by a small number of k-shells, this strategy is even less
efficient than the random strategy for some values of $q$, while the
improved robustness for degree and betweenness strategies is evident
compared with the random selection.  While in the random case, for
$q\gtrsim0.4$, a significant decrease of the robustness with $q$ is
observed, in the degree and betweenness cases, the change is smoother
and only significantly drops for higher values of $q$.  A maximum in the
ratio $R/R_\mathrm{random}$ occurs for $q\approx0.85$, where the
relative improvement is above $12\%$. 
Since, in random networks, many metrics are strongly
correlated \cite{Newman10}, the results for betweenness and degree are similar.

 Many real-world systems are characterized by a degree distribution
which is scale free with a degree exponent $\gamma$
\cite{Albert00,Clauset09}.  In Fig.~\ref{fig::robustness}(b) we plot $R$
as a function of $q$ for two coupled scale-free networks (SF) with
$10^3$ nodes each and $\gamma=2.5$.  Similar to the two coupled ER, this
system is also significantly more resilient when the autonomous nodes
are selected according to the highest degree or betweenness.  For
values of $q\lesssim0.85$ the robustness is similar to that of a single
network ($q=0$) since the most relevant nodes are decoupled.  A peak in
the relative robustness, $R/R_\mathrm{random}$ (see inset of
Fig.~\ref{fig::robustness}b), occurs for $q\approx0.95$ where the
improvement, compared to the random case, is almost $30\%$.
Betweenness, degree, and $k$-shell, have similar impact on the
robustness since these three properties are strongly correlated for SF.
From Fig.~\ref{fig::robustness}, we see that, for both SF and ER, the
robustness is significantly improved by decoupling, based on the
betweenness, less than $15\%$ of the nodes.  Studying the dependence of
the robustness on the average degree of the nodes we conclude that for average
degree larger than five, even $5\%$ autonomous nodes are enough to
achieve more than $50\%$ of the maximum possible improvement.

    \begin{figure}
      \includegraphics[width=0.7\columnwidth]{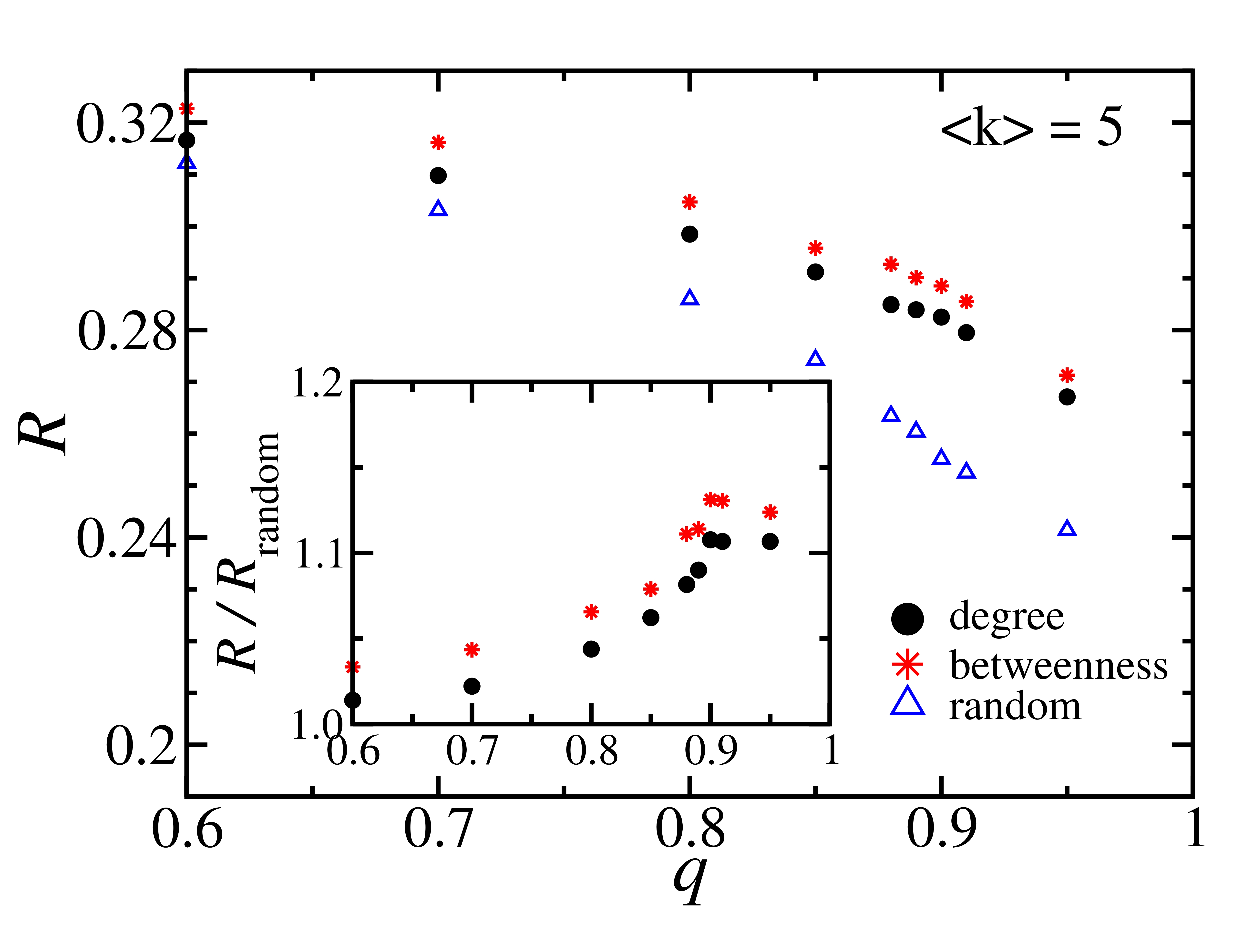} \\
      \caption{
        \label{fig::fourER}
	Dependence of the robustness, $R$, on the degree of coupling,
$q$, for two, randomly interconnected modular networks with $2\cdot10^3$
nodes each.  The modular networks were obtained from four
Erd\H{o}s-R\'enyi networks, with $500$ nodes each and average degree
five, by randomly connecting each pair of modules with an additional
link.  Autonomous nodes are selected in three different ways: randomly
(blue triangles), higher degree (black dots), and higher betweenness
(red stars).  In the inset we see the relative enhancement of the
robustness, for the second and third schemes of autonomous selection
compared with the random case.  Results have been averaged over $10^2$
configurations and $10^3$ sequences of random attacks to each one.
      }
    \end{figure}
For the cases discussed in Fig.~\ref{fig::robustness},
results obtained by selecting autonomous nodes based on the highest
degree do not significantly differ from the ones based on the highest
betweenness.  This is due to the well known finding that for
Erd\H{o}s-R\'enyi and scale-free networks, the degree of a node is
strongly correlated with its betweenness \cite{Newman10}.  However, many
real networks are modular, i.e., composed of several different modules
interconnected by less links, and then nodes with higher betweenness are
not, necessarily, the ones with the largest degree \cite{Cohen10}.
Modularity can be found, for example, in metabolic systems, neural
networks, social networks, or infrastructures
\cite{Ravasz02,Happel94,Gonzalez07,Eriksen03}.  In
Fig.~\ref{fig::fourER} we plot the robustness for two coupled modular
networks.  Each modular network was generated from a set of four
Erd\H{o}s-R\'enyi networks, of $500$ nodes each and average degree five,
where an additional link was randomly included between each pair of
modules.  For a modular network, the nodes with higher betweenness are
not necessarily the high-degree nodes but the ones bridging the
different modules.  Figure~\ref{fig::fourER} shows that the strategy
based on the betweenness emerges as better compared to the high degree
method.

    \begin{figure}
      \includegraphics[width=0.7\columnwidth]{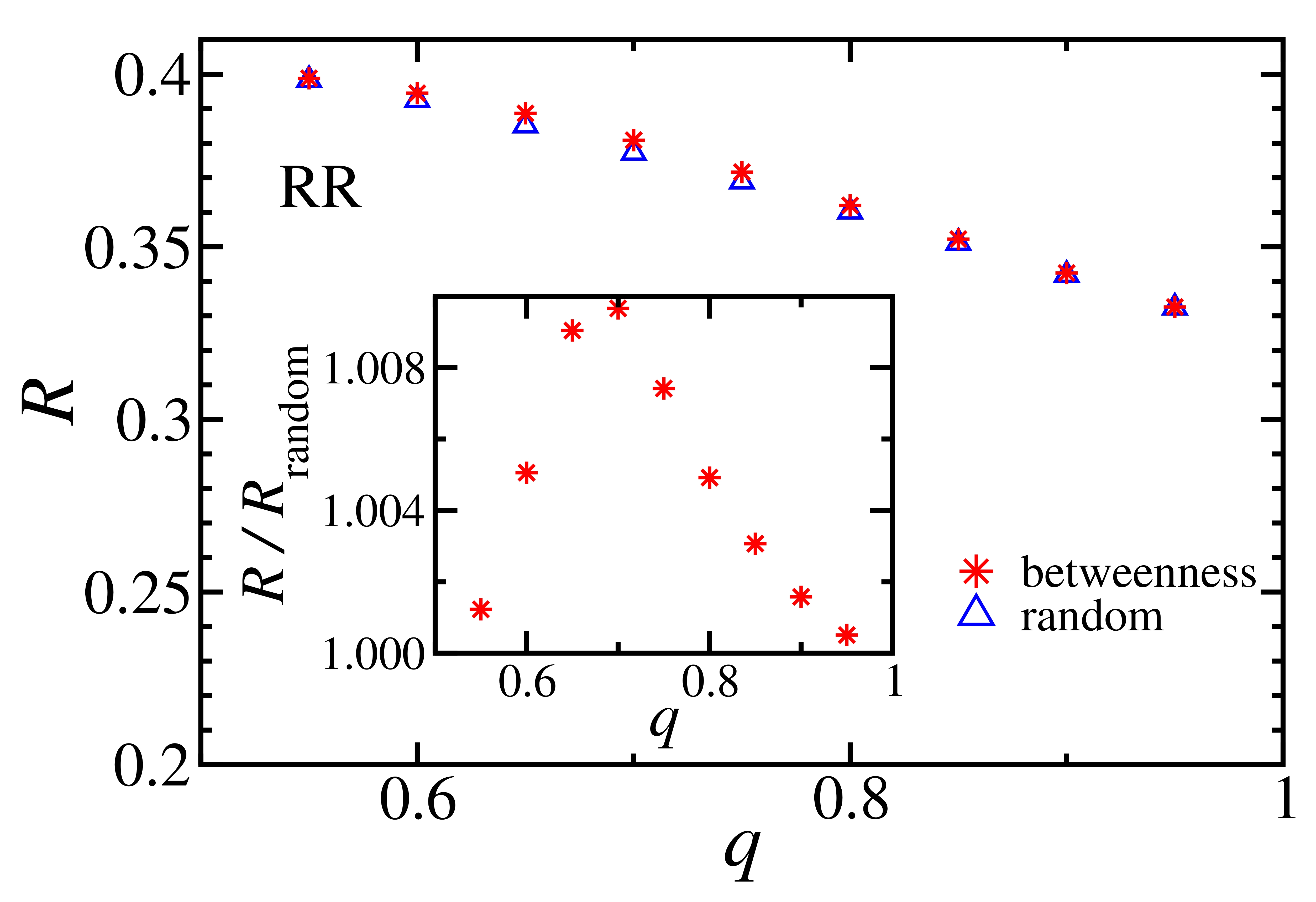}
      \caption{
        \label{fig::RR}
	Dependence of the robustness, $R$, on the degree of coupling,
$q$, for two, randomly interconnected random regular graphs with
$8\cdot10^3$ nodes each, all with degree four.  Autonomous nodes are
selected in two different ways: randomly (blue triangles) and higher
betweenness (red stars).  In the inset the relative enhancement of the
robustness is shown for the betweenness compared to the random case.
Results have been averaged over $10^2$ configurations and $10^3$
sequences of random attacks to each one.
      }
    \end{figure}
  Another example that shows that betweenness is superior to degree is
when we study coupled random regular graphs.  In random regular graphs
all nodes have the same degree and are connected randomly.
Figure~\ref{fig::RR} shows the dependence of the robustness on the
degree of coupling, for two interconnected random regular graphs with
degree $4$.  The autonomous nodes are selected randomly (since all
degrees are the same) or following the betweenness strategy.  Though all
nodes have the same degree and the betweenness distribution is narrow,
selecting autonomous nodes based on the betweenness is always more
efficient than the random selection.  Thus, the above two examples
suggest that betweenness is a superior method to chose the autonomous
nodes compared to degree.

\begin{figure}
  \includegraphics[width=0.7\columnwidth]{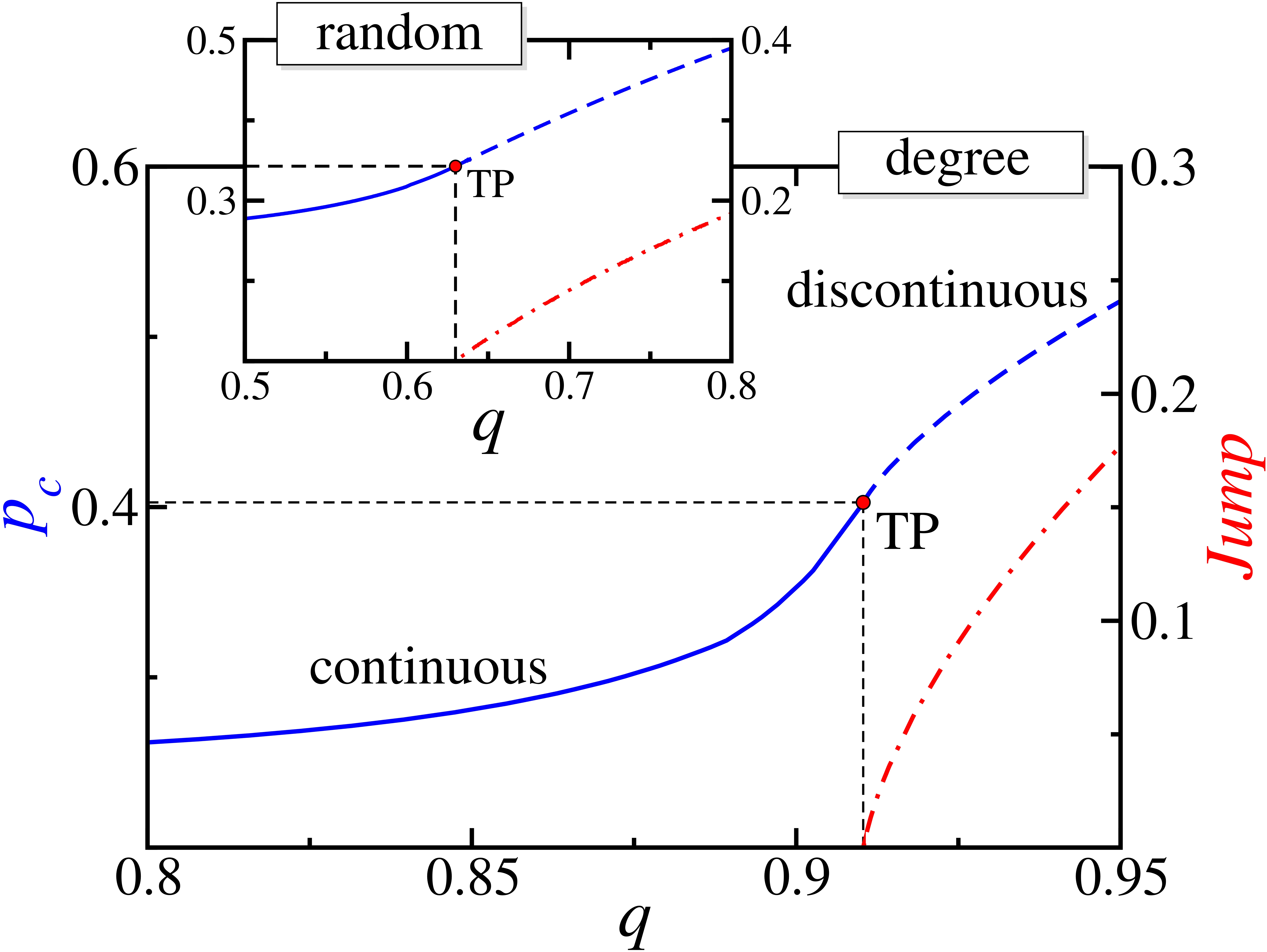}
  \caption{
    Two-parameter diagram (blue curves) of two coupled ER
(average degree $\langle k\rangle = 4$) under random attack. The horizontal axis is the
degree of coupling $q$ and the vertical one is $p$ so that $1-p$ is
the fraction of initially removed nodes.  The size of the jump in the fraction of
$A$-nodes in the largest connected cluster is also included
(\mbox{red-dotted-dashed} curve).  The dashed curve stands for a
discontinuous transition while the solid one is a critical line
(continuous transition). The two lines meet at a tricritical point (TP).
Autonomous nodes are selected based on the degree (main plot) and
randomly (inset). Results have been obtained with the formalism of
generating functions. 
    \label{fig::diagram}
  }
\end{figure}
\begin{figure}
  \includegraphics[width=0.7\columnwidth]{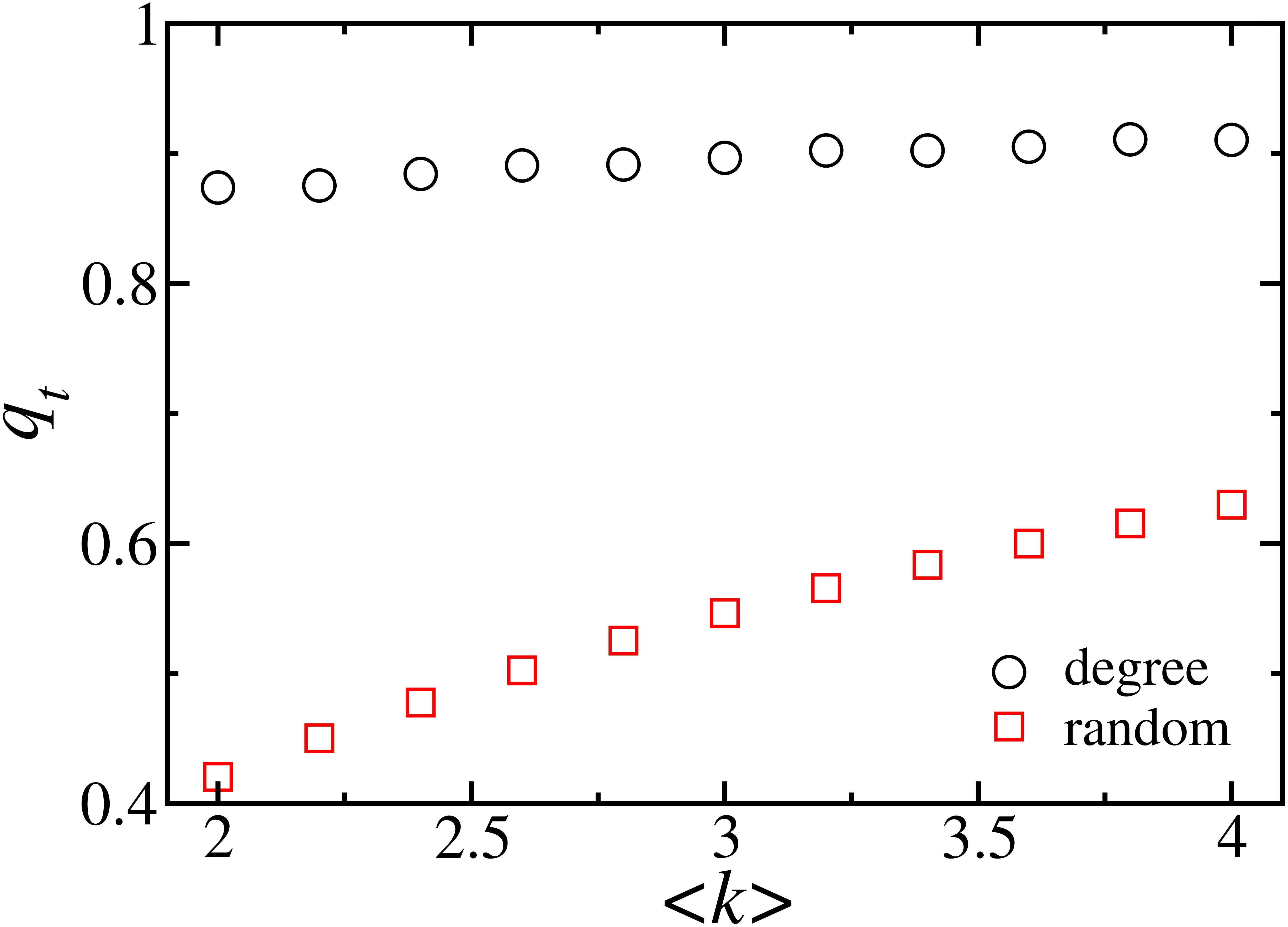}
  \caption{
    Tricritical coupling $q_t$ dependence on the average
degree $\langle k\rangle$ for two coupled ER, showing that the fraction of autonomous nodes
to smoothen out the transition is significantly reduced with the proposed
strategy when compared with the random case. Autonomous nodes are
selected following two different strategies: randomly (red squares) and
high degree (black circles).
    \label{fig::qtk}
  }
\end{figure}
  The vulnerability is strongly related to the degree of coupling $q$.
Parshani {\it et al.} \cite{Parshani10} have analytically and
numerically shown that, for random coupling,  at a critical
coupling $q=q_t$, the transition changes from continuous (for $q<q_t$)
to discontinuous (for $q>q_t$). In Fig.~\ref{fig::diagram} we see the
two-parameter diagram ($p_c$ vs $q$) with the tricritical point and the transition
lines (continuous and discontinuous) for the random (inset) and the
degree (main) strategies.  As seen in Fig.~\ref{fig::diagram}, when autonomous nodes are randomly selected,
about $40\%$ autonomous nodes are required to soften the transition and
avoid catastrophic cascades, while following the strategy proposed here
only a relatively small amount ($q>0.9$) of autonomous nodes are needed
to avoid a discontinuous collapse. Above the tricritical point, the jump
increases with the degree of coupling, lending arguments to the paramount
necessity of an efficient strategy for autonomous selection, given that
the fraction of nodes which can be decoupled is typically limited. The
dependence of $q_t$ on the average degree $\langle k\rangle$ is shown in
Fig.~\ref{fig::qtk}. The ratio between the tricritical coupling for
degree and random strategies increases with decreasing $\langle k\rangle$.
For example, for \mbox{$\langle k\rangle \approx 2$} the fraction of
autonomous nodes needed to soften the transition with the random selection is
six times the one for the degree strategy.

\begin{figure}
  \includegraphics[width=0.7\columnwidth]{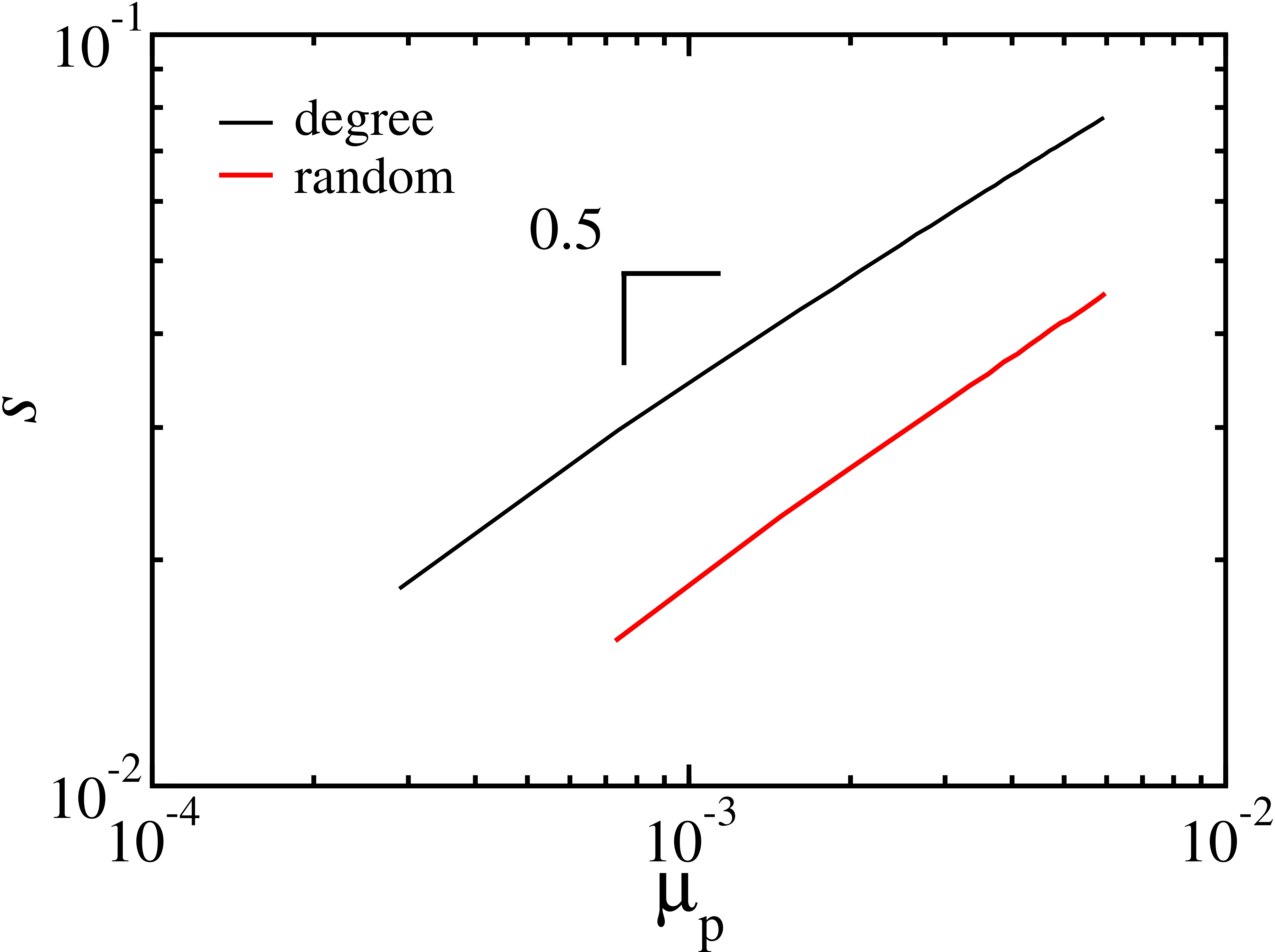}
  \caption{ 
    Dependence of the fraction of $A$-nodes in the largest connected
cluster on the scaling field $\mu_p$ along the direction perpendicular
to the transition line at the tricritical point. The slope is the
tricritical exponent $\beta_t$ related with the order parameter.
Autonomous nodes in the two coupled ER \mbox{($\langle k\rangle =4$)} have been selected
randomly (red line) and following the ranking of degree (black line).
    \label{fig::crossover}
  }
\end{figure}
As in Ref.~\cite{Araujo11}, following the theory of Riedel and Wegner
\cite{Riedel69,Riedel72,Riedel72b}, we can characterize the
tricritical point. Two relevant scaling fields are defined: one tangent
($\mu_p$) and the other perpendicular ($\mu_q$) to the critical curve at
the tricritical point.  In these coordinate axes the continuous line is
described by $\mu_p\sim\mu_q^{1/\varphi_t}$, where the tricritical
crossover exponent $\varphi_t=1.00\pm0.05$ for degree and random
strategies.  The tricritical order parameter exponent, $\beta_t$, can be
evaluated from,
\begin{equation}\label{eq::tricritial_beta}
s\left(\mu_p,\mu_q=0\right)\sim\mu_p^{\beta_t} ,
\end{equation}
giving $\beta_t=0.5\pm0.1$ for both strategies. Since these two
exponents are strategy independent (see Fig.~\ref{fig::crossover}), we conjecture that the tricritical
point for degree and random selection are in the same universality
class.

\section*{Discussion}

  Here, we propose a method to chose the autonomous nodes in order
to optimize the robustness of coupled networks to failures.  We find the
betweenness and the degree to be the key parameters for the selection of
such nodes and we disclose the former as the most effective for modular
networks. Considering the real case of the Italian communication network
coupled with the power grid, we show in Fig.~\ref{fig::italy} that
protecting only the four communication servers with highest betweenness
reduces the chances of catastrophic failures like that witnessed during the
blackout in 2003.  When this strategy is implemented the resilience to
random failures or attacks is significantly improved and the fraction of
autonomous nodes necessary to change the nature of the percolation
transition, from discontinuous to continuous, is significantly reduced.
We also show that, even for networks with a narrow distribution of node
degree like Erd\H{o}s-R\'enyi graphs, the robustness can be significantly
improved by properly choosing a small fraction of nodes to be autonomous.
As a follow-up it would be interesting to understand how correlation
between nodes, as well as dynamic processes on the network, can
influence the selection of autonomous nodes.  Besides, the cascade
phenomena and the mitigation of vulnerabilities on regular lattices
and geographically embedded networks are still open
questions. It is important to note that while we use here high
betweenness and high degree as a criterion for autonomous nodes, it is
possible that other metrics will be also useful. For example, the
eigenvector component of the largest eigenvector of the adjacency
matrix (even weighted) makes a very good candidate (see
e.g.~\cite{vanMieghem11}).

\section*{Methods}

We consider two coupled networks, $A$ and $B$, where a fraction of $1-p$
$A$-nodes fails. The cascade of failures can be described by the
iterative equations, Eqs.~(\ref{eq::iterative})
\cite{Buldyrev10,Parshani10}, where $\alpha_n$ and $\beta_n$ are,
respectively, the fraction of $A$ and $B$ surviving nodes at iteration
step $n$ (not necessarily in the largest component), and
$S_{x}(\chi_n)$ ($\chi = \alpha | \beta$, $x = A | B$) is the
fraction of nodes in the largest component in network $x$ given that $1
- \chi$ nodes have failed. This can be calculated for coupled networks
  in the thermodynamic limit ($N\rightarrow\infty$) using generating
functions.

\subsection*{Random Protection}

As proposed by Parshani {\it et al.} \cite{Parshani10}, when autonomous
nodes are randomly selected and the degree of coupling is the same in $A$
and $B$, the set of Eqs.~(\ref{eq::iterative})
simplifies to
\begin{eqnarray}
\alpha_1 &=& p , \nonumber \\
\beta_n &=& 1 - q \left[ 1 - S_A(\alpha_n) p \right] , \\
\alpha_n &=& p \left( 1 - q \left[ 1 - S_B(\beta_{n-1}) \right]
\right) , \nonumber
\end{eqnarray}
where $q$ is the degree of coupling. 
The degree distribution of the networks does not change
in the case of random failures and $S_{x}(\chi_n)$ can be calculated as
  \mbox{$S_x(\chi) = 1 - G_{P_x}\left( 1-\chi(1-u_x)
\right)$}~\cite{Newman02}, 
where $G_{P_x}(z)$ is the generating function of the degree distribution
of network $x$,
\begin{equation}
  G_{P_x}(z) = \sum_k {P_x(k)z^k} ,  \nonumber
\end{equation}
and $u_x$ satisfies the transcendental equation
\begin{equation}
u_x = \frac {G'_{P_x}\left(1 - \chi(1-u_x)\right)} {G'_{P_x}(1)}. \nonumber
\end{equation}

The size of the largest component in network $x$ is given by
$\chi S_{x}(\chi)$.
 
For ER networks $G_{P_x}(y)=\exp\left[\left< k \right>_x(z-1)\right]$,
where $\left< k \right>_x$ is the average number of links in network
$x$, and therefore

\begin{equation}
\frac {G'_{P_x}\left(y\right)} {G'_{P_x}(1)} = G_{P_x}(y) . \nonumber
\end{equation}
With the above equations one can calculate the size of the largest
component in both networks at the end of the cascade process.

Recently, Son {\it et al.} \cite{Son11b} proposed an equivalent scheme
based on epidemic spreading to solve the random protection case.

\subsection*{High Degree Protection}
When autonomous nodes are selected following the degree strategy, the
fraction of dependent nodes $q_{xn}$ changes with the iteration step $n$
and the set of Eqs.~\ref{eq::iterative} no longer simplifies. We divide the
discussion below into three different parts: the degree distribution,
the largest component, and the coupling (fraction of dependent nodes).

\subsubsection*{The Degree Distribution}
The networks A and B are characterized by their degree distributions,
$P_A(k)$ and $P_B(k)$, which are not necessarily the same. The developed
formalism applies to any arbitrary degree distribution. We start by
first splitting the degree distribution into two parts, the component
corresponding to the low-degree dependent nodes, $P_{xD}(k)$, and the
component corresponding to the high-degree autonomous ones,
$P_{xI}(k)$. To accomplish this, one must determine two parameters,
the maximum degree of dependent nodes, $k_{xm}$, and the fraction of
nodes with degree $k_{xm}$ that are coupled with the other network,
$f_{xm}$. These two parameters can be obtained from the relations,
\begin{equation}
  \sum_{k=0}^{k_{xm}-1} {P_x(k)} < q < \sum_{k=0}^{k_{xm}} {P_x(k)}
\nonumber
\end{equation}
and
\begin{equation}
  \sum_{k=0}^{k_{xm}-1} {P_x(k)} + f_{xm} P_x(k_{xm}) = q , \nonumber
\end{equation}
where $q_{x}$ is the initial degree of coupling.
One can then
write
\begin{equation}
  P_{xD}(k) =  \left\{ \begin{array}{ll}
  P_x(k), & k<k_{xm} \\
  f_{xm}P_x(k), & k=k_{xm} \\
  0, & k>k_{xm}
\end{array} \right. 
\end{equation}
and
\begin{equation}
  P_{xI}(k) =  \left\{ \begin{array}{ll}
  0, & k<k_{xm} \\
  (1-f_{xm})P_x(k), & k=k_{xm} \\
  P_x(k), & k>k_{xm}
\end{array}. \right.
\end{equation}

In the model, a fraction of $1-p$ $A$-nodes are randomly removed. If, at
iteration step $n$, $\alpha_n$ nodes survive \mbox{($\alpha_n\leq p$)},
$p(1-q_A)$ nodes are necessarily autonomous and the remaining ones,
$\alpha_n-p(1-q_A)$, are dependent nodes. One can then show
that the degree distribution of network $A$, under the failure of
$1-\alpha_n$ nodes, $P'_{A,n}(k)$, is given by
\begin{equation}
  P_{A,n}'(k) = \left(\frac{1- \frac{p}{\alpha_n}(1-q)}{q}\right) 
P_{AD}(k) + \frac{p}{\alpha_n} P_{AI}(k), \nonumber
\end{equation} 
while the fraction of surviving links is
\begin{equation}
  p_{An} = \alpha_n \frac{\sum_k {k P_{A,n}'(k)}}{\sum_k {k P_{A}(k)}} .
\nonumber
\end{equation}

All the $B$-nodes which do not survive are dependent and so
the degree distribution at iteration $n$,
$P_{B,n}'(k)$, is given by
\begin{equation}
  P_{B,n}'(k) = \left(\frac{1-\frac{1}{\beta_n}(1-q)}{q}\right) P_{BD}(k)
+ \frac{1}{\beta_n}P_{BI}(k), \nonumber
\end{equation}
while the fraction of surviving links is
\begin{equation}
  p_{Bn} = \beta_n \frac{\sum_k {k P_{B,n}'(k)}}{\sum_k {k P_{B}(k)}}
\nonumber \ \ .
\end{equation}

\subsubsection*{The Largest Component}
With the degree distribution $P_{x,n}'(k)$ and the fraction of surviving
links $p_{xn}$ one can calculate the size of the largest component as
\begin{eqnarray}
  S_{x}(\chi_n) &=&  1 - G_{P'_{xn}}\left( 1-p_{xn} +p_{xn}
\tau_{xn}\right) \nonumber\\
   &=& 1 - \sum_k {P_{xn}'(k) \left( 1-p_{xn} +p_{xn}
\tau_{xn}\right)^k}, \nonumber
\end{eqnarray}
where $\tau_{xn}$ satisfies the self consistent equation
\begin{eqnarray}
  \tau_{xn} &=& \frac{G'_{P'_{xn}}\left( 1-p_{xn} +p_{xn} \tau_{xn}\right)} {G'_{P'_{xn}}(1)} \nonumber \\
         &=& \frac{\sum_k{k P_{xn}'(k) \left( 1-p_{xn} +p_{xn}
\tau_{xn}\right)^{k-1}}} {\sum_k{k P_{xn}'(k)}} .\nonumber
\end{eqnarray}

\subsubsection*{The coupling}
To calculate the fraction $q_{\alpha,n}$ (and $q_{\beta,n}$)
one must first calculate the degree distribution of the nodes in the
largest component. This is given by
\begin{equation}
  P_{xG,n}(k) = P_{xn}'(k) \frac{1 -\left( 1-p_{xn} +p_{xn}
\tau_{xn}\right)^k } {S_{\chi_nx}(\chi_n)}.\nonumber
\end{equation}
The fraction of nodes in the largest component that are autonomous is then
given by
\begin{equation}
  q_{xG,n} = \left(1-f_{xm}\right) P_{x G,n}(k_{xm}) + \sum_{k=k_{xm}+1}
^{\infty} { P_{xG,n}(k)} ,\nonumber
\end{equation}
where the upper limit of the sum is the maximum degree in the network,
which we consider to be infinity in the thermodynamic limit. The fraction of autonomous nodes from the original network
remaining in the largest component is
  \mbox{$q_{xG,n}\chi_n S_{x}(\chi_n)$},
while the total fraction of autonomous nodes is given by
  \mbox{$1 - q$}.
The fraction of nodes disconnected from the largest component that are
autonomous is then given by
\begin{equation}
  1-q_{x,n} = \frac{1 - q - q_{xG,n}\chi_n S_{x}(\chi_n)} {1 -\chi_n
S_{x}(\chi_n) } , \nonumber
\end{equation}
so that the fraction of dependent nodes which have fragmented from the
largest component is
\begin{equation}
  q_{x,n} = 1 - \frac{1 - q - q_{xG,n}\chi_n S_{x}(\chi_n)} {1 -\chi_n
S_{x}(\chi_n) } . \nonumber
\end{equation}

For simplicity, here we assume that $k_{x,m}$ and $f_{xm}$ are
constant and do not change during the iterative process. In fact, this
is an approximation as the degree of the autonomous nodes is expected
to change when their neighbors fail. However, in spite of shifting the
transition point, this consideration does not change the global picture
described here.

\subsection*{Numerical simulations}

Numerical results have been obtained with the efficient algorithm
described in Ref.~\cite{Schneider13} for coupled networks.

\begin{acknowledgments}
  We acknowledge financial support from the ETH Risk Center, from the
Swiss National Science Foundation (Grant No. 200021-126853), and the
grant number FP7-319968 of the European Research Council.  We thank the
Brazilian agencies CNPq, CAPES and FUNCAP, and the grant CNPq/FUNCAP. SH
acknowledges the European EPIWORK project, the Israel Science
Foundation, ONR, DFG, and DTRA.
\end{acknowledgments}

\end{document}